\documentclass[10pt]{wlscirep}
\usepackage{braket}
\usepackage[utf8]{inputenc}
\usepackage[T1]{fontenc}
\usepackage{subcaption}
\usepackage{amsthm, latexsym, amsmath, amssymb}
\usepackage{enumitem}
\usepackage{float}
\usepackage{xcolor}
\usepackage{soul}
\usepackage{comment}
\usepackage{listings}
\usepackage{placeins}

\usepackage{physics}
\usepackage{hyperref}

\DeclareMathOperator*{\argmax}{argmax}
\DeclareMathOperator{\dist}{dist}

\newcommand{\Cost}{\mathrm{C}}

\renewcommand{\vec}{\vb*}

\title{Quantum Approximate Optimization Algorithms for Molecular Docking}

\author[2,*]{Christos Papalitsas} 
\author[3,*]{Yanfei Guan}
\author[1,+]{Shreyas Waghe}
\author[2,+]{Athanasios Liakos}
\author[2,+]{Ioannis Balatsos}
\author[1,+]{Vassilios Pantazopoulos}

\affil[1]{Pfizer AIDA, Scientific Computing \& HPC, Cambridge MA 02139, USA}
\affil[2]{Pfizer AIDA, Scientific Computing \& HPC, Thessaloniki 55535, Greece}
\affil[3]{Pfizer Worldwide Research, Development \& Medical, Cambridge MA 02139, USA}

\affil[*]{christos.papalitsas@pfizer.com, yanfei.guan@pfizer.com}

\affil[+]{these authors contributed equally to this work}

\begin{abstract}

Molecular docking is a critical process for drug discovery and challenging due to the complexity and size of biomolecular systems, where the optimal binding configuration of a drug to a target protein is determined. Hybrid classical-quantum computing techniques offer a novel approach to address these challenges. The Quantum Approximate Optimization Algorithm (QAOA) and its variations are hybrid classical-quantum techniques, and a promising tool for combinatorial optimization challenges. This paper presents a Digitized Counterdiabatic QAOA (DC-QAOA) approach to molecular docking. Simulated quantum runs were conducted on a GPU cluster. We examined 14 and 17 nodes instances - to the best of our knowledge the biggest published instance is 12-node at Ding et al.\cite{ding2024molecular},  and we present the results.
Based on computational results, we conclude that binding interactions represent the anticipated exact solution. Additionally, as the size of the examined instance increases, the computational times exhibit a significant escalation.

\end{abstract}
\begin{document}

\flushbottom
\maketitle
%
%
\thispagestyle{empty}



\section*{Introduction} 
\label{intro}

Drug development is a complex and costly process requiring extensive research. Among structure-based drug design methods, molecular docking is commonly used to predict binding modes and affinities between drug molecules and target proteins.\cite{BELFIELD2023100251}

Quantum computing has arisen as an auspicious technology for solving complex computational problems, particularly in the territory of  combinatorial optimization and simulation.\cite{feynman1998feynman} \cite{feynman_1982} Among these, molecular docking -a crucial process in drug discovery that demands the prediction of ligand binding to protein targets, has drawn significant interest due to its high computational needs. While quantum processes such as quantum annealing (QA) and the variational quantum eigensolver (VQE) show potential for accelerating docking simulations, the current limitations in quantum hardware constrain to hybrid approaches. One such approach leverages classical hardware accelerators, particularly graphics processing units (GPUs), to simulate quantum algorithms efficiently. This study explores the implementation of simulated quantum runs for molecular docking on GPUs, focusing on the potential for scalability, speedup, and enhanced accuracy. 

Recent advancements of quantum computing technologies has brought new capabilities to various scientific disciplines, particularly those involving complex optimization problems. One field that stands to benefit significantly from these advancements is computational biology, where the intricate process of drug discovery often involves solving challenging combinatorial optimization tasks. Among these tasks, molecular docking is of paramount importance, as it involves predicting the most favorable binding configurations between small molecules (such as potential drugs) and their target proteins. The accuracy and efficiency of these predictions are critical to the success of drug discovery efforts. Traditionally, classical algorithms have been employed to address molecular docking challenges; however, these algorithms face significant limitations in terms of scalability and computational efficiency when dealing with large and complex systems. \cite{dupont2023quantum}

The convergence of quantum computing and computational biology offers opportunities for advancing drug discovery processes. One such advancement is the application of the quantum approximate optimization algorithm (QAOA) to molecular docking—a critical task in computational biology that seeks to predict the optimal binding configurations between small molecules and target proteins. As drug discovery continues to rely heavily on the accuracy and efficiency of molecular docking, there is a growing interest in exploring quantum computing techniques like QAOA to overcome the limitations of classical algorithms. QAOA has gained significant attention in recent years due to its potential to solve combinatorial optimization problems. Among the various applications, the max clique problem stands out as a benchmark problem for evaluating QAOA’s performance.  \cite{tomesh2022quantum,hadfield2019quantum}

QAOA is a hybrid classical-quantum algorithm designed to tackle combinatorial optimization problems, which are often hard to be solved for classical computers alone. The algorithm operates by leveraging the quantum properties of superposition and entanglement to explore multiple possible solutions simultaneously, thereby increasing the likelihood of finding an optimal or near-optimal solution. One of the most compelling aspects of QAOA is its ability to approximate solutions to NP-hard problems, which are known for their computational complexity. Among the various combinatorial problems, the max clique problem has emerged as a critical benchmark for evaluating the performance of QAOA. \cite{chandarana2022digitized,hadfield2019quantum}

The max clique problem involves finding the largest subset of vertices in a graph, where every two vertices are connected. It is a well-known NP-hard problem that has direct implications for many optimization tasks, including those found in molecular docking. The ability of QAOA to approximate solutions to such challenging problems with quantum advantage has spurred interest in its application to molecular docking scenarios.\cite{seda2023maximum}

The k-clique problem is a problem of identifying completely connected sub-graphs or cliques within a graph, each consisting of exactly k vertices. The highest weight k-clique problem extends the concept to vertex-weighted graphs and its goal is to find the k-clique with the highest sum of vertex weights within the graph. Another challenging problem is the N highest weight k-cliques problem, where the goal is to find a set of up to N weighted cliques whose weights are among the highest of all k-cliques in a given graph. In cases where the graph contains less than N k-cliques, the algorithm returns all k-cliques within a graph. Several algorithms have been developed to find k-cliques in unweighted graphs and also to detect highest weight k-cliques in vertex-weighted graph. 

In molecular docking, the problem of identifying the optimal binding configuration between a drug and its target protein can be mapped onto an optimization problem, where the goal is to find the configuration that minimizes the binding energy. This optimization task can be viewed as a combinatorial problem, where various possible configurations must be evaluated to determine the best one. Classical algorithms, such as genetic algorithms, monte carlo simulations, and molecular dynamics, have traditionally been used to solve this problem. However, these methods often require significant computational resources and may struggle to find the global optimum in large and complex systems. This is where QAOA offers a potential advantage. By framing the molecular docking problem as a combinatorial optimization task, QAOA can be employed to explore the solution space more efficiently, potentially leading to better docking predictions and faster drug discovery.\cite{rozman2024enhanced}

The potential of QAOA to enhance molecular docking is particularly relevant in the current era of quantum computing, known as the noisy intermediate-scale quantum (NISQ) era. NISQ devices, while not yet capable of fault-tolerant quantum computation, offer a limited number of qubits and are subject to noise and decoherence. Despite these limitations, NISQ devices have demonstrated the potential to achieve quantum advantage for specific tasks when used in conjunction with classical computing resources. Hybrid quantum-classical algorithms like QAOA are well-suited to this environment, as they allow for the distribution of computational tasks between quantum and classical processors. In the case of molecular docking, the quantum processor can be used to perform the most challenging parts of the optimization, while the classical processor handles less complex tasks. This hybrid approach not only maximizes the utility of current quantum devices but also provides a practical pathway for integrating quantum computing into existing computational biology workflows. \cite{chen2024noise}

A critical consideration in the practical deployment of QAOA for molecular docking is the choice of computing platforms. Given the intensive computational demands of molecular docking, it is essential to utilize hardware that can efficiently manage large-scale calculations. Graphics processing units (GPUs) have emerged as a powerful tool in this context, offering high degrees of parallelism and computational power. Originally designed for rendering graphics, GPUs have found widespread application in scientific computing due to their ability to perform many calculations simultaneously.

GPUs have already demonstrated their utility in classical molecular docking algorithms, and their integration with quantum algorithms like QAOA could further accelerate drug discovery processes. At this work we implement QAOA on GPUs to leverage their computational advantages. To enhance the performance of QAOA, particularly when running on GPUs, warm-starting techniques have been explored. Warm-starting involves initializing the quantum algorithm with a solution obtained from a classical algorithm or a previous iteration, potentially reducing the number of quantum operations required to reach an optimal solution. This technique is especially useful in the NISQ era, where minimizing quantum operations is crucial due to noise and decoherence in quantum devices. By combining warm-starting with GPU acceleration, we aim to demonstrate significant improvements in both accuracy and efficiency for molecular docking tasks. \cite{egger2021warm}

Through a series of experiments, we explore the parameters that influence QAOA’s performance, the scalability of the algorithm on GPUs, and its integration with existing molecular docking frameworks. Our findings suggest that QAOA, when optimized for GPU architectures and enhanced with techniques like warm-starting, can significantly speed up the molecular docking process, offering a promising avenue for accelerating drug discovery pipelines in the near term.

Warm-starting technique is being used to enhance the performance of QAOA when applied to molecular docking. Warm-starting involves initializing the quantum algorithm with a solution obtained from a classical algorithm or from a previous iteration of the quantum algorithm. This strategy can significantly reduce the number of quantum operations required to reach an optimal solution, making the algorithm more efficient and less susceptible to the effects of noise and decoherence. In the context of the NISQ era, where quantum resources are limited, warm-starting represents a valuable tool for maximizing the effectiveness of QAOA. Our experiments demonstrate that warm-starting, when combined with the parallel processing capabilities of GPUs, can lead to substantial improvements in both the accuracy and efficiency of molecular docking predictions. \cite{egger2021warm}

Recent advancements in approximate optimization algorithms, such as quantum approximate optimization algorithms (QAOA) \cite{ding2024molecular,medina2024recursive, chen2024noise }, quantum annealing approaches \cite{zha2023encoding, li2024efficient}, bio-inspired \cite{garcia2019bio} and tabu search algorithms \cite{gendreau1993solving}, have shown promising applications in molecular docking, enabling efficient exploration of binding conformations and energy landscapes \cite{gendreau1993solving, fang2014solving,li2010efficient,thomsen2006moldock,zha2023encoding}.

This paper is structured as follows: Chapter \hyperref[intro]{1} provides an introduction to the topic, outlining the motivation for exploring QAOA in the context of molecular docking and the potential benefits of using GPU platforms. In Chapter \hyperref[methods]{2}, we present the methods used in our research, including the experimental setup, the specific QAOA implementation, and the GPU configurations employed. In Chapter \hyperref[results]{3}, we present the computational results of our experiments, offering an in-depth analysis of the performance of QAOA on GPU platforms. 
Chapter \hyperref[conclusion]{4} is a discussion of our findings and practical applications, including potential avenues for future work and further improving the efficiency and scalability of QAOA in molecular docking. Through this comprehensive exploration, we aim to contribute to the ongoing efforts to harness the power of quantum computing for advancing drug discovery and other critical areas of computational biology.

\section*{Methods}
\label{methods}

Molecular docking for small molecules is a commonly used computational chemistry tool in structure-based drug design. Throughout drug discovery, when ligand-protein complex structures are available through experimental methods or modeling, molecular docking are widely used for SAR (structure activity relationship) interpretation, molecular optimization, and virtual screening.\cite{pagadala2017software} Molecular docking methods usually contains two steps: generating the docking pose of suitable conformation, positions, and orientations; assessing the pose with scoring functions. In conventional docking methods, the pose generation process is mostly through searching or sampling algorithms. Due to the size of target proteins, the degree of freedom is usually not manageable. In most practice of molecular docking, ligands position, orientation, and conformations are sampled through systematic or stochastic methods.\cite{STANZIONE2021273} Although the protein pocket remains fixed during the sampling, the large degree of freedom still makes it challenging to cover the space or consume resources and time to find a reasonable pose. In order to further simplify the complexity from degree of freedom, the binding pocket and ligands can be represented as pharmacophores (phc4s) which define critical non-covalent interactions between ligand and protein. These pharmacophores are set of points with radius abstracted from key residues and functional groups of protein pocket and ligand. In a ligand-protein binding complex, ligand pharmacophores need to align with the pocket to allow favorable interactions. 

With the ligand molecule and binding site abstracted as a set of pharmacophore points, the molecular docking problem can be redefined as finding the maximum degree of contact between those points, which thus define the translation and rotation of the ligand and protein for the optimum binding. Pioneered by Kuhl {\it et al.}, maximizing the degree of pharmacophore contacts were tackled by clique-detecting methods in graph theory.\cite{kuhl1984combinatorial}

\subsection*{Combinatorial Optimization Background}

From the combinatorial lens, at the heart of molecular docking is an assignment problem (AP), where each pharmacophore on either chemical species must be assigned at most one pharmacophore from the other species. The assignment problem in its generality is known to be NP-Complete, i.e., given a candidate solution to the problem, it is possible to verify its correctness efficiently, but there is no known efficient algorithm to find such solutions. The assignment problem is ubiquitous across disciplines, and common examples include job scheduling, where tasks need to be assigned to workers in a way that minimizes total completion time; resource allocation, where resources must be distributed among various projects to maximize efficiency; and matching problems, such as pairing students with schools or matching organ donors with recipients. This problem also appears in transportation logistics, where goods need to be assigned to delivery routes to minimize costs, and in network design, where connections between nodes must be optimized for performance and reliability.
 \cite{kelley2010optimization}


Through manual expertise of the chemical instance, it is possible to limit the number of possible permissible assignments to obtain computationally tractable instances of AP. In contrast, we seek a general method of analysis for molecular docking, which does away with this manual filtering of assignment combinations. An amenable method towards this goal is formulating molecular docking as a maximum vertex-weight clique problem (MVWCP) on a particular graph encoding of the chemical regime called the binding interaction graph or docking graph. We briefly review some prerequisite combinatorial optimization knowledge in this section. \cite{gendreau1993solving}

A combinatorial optimization problem over a bitstring $z \in \{0,1\}^N$ is of the form $z^* = \argmax_z \Cost(z)$ where $z$ comprises a collection of \textit{clauses} $z_\alpha$. Each clause $z_\alpha$ relates a number of elements of the bitstring, and the objective function of the problem can be decomposed as $\Cost(z) = \sum_{\alpha} \Cost(z_\alpha)$. In our particular instance, the combinatorial optimization problem is over a simple undirected weighted graph. 
A simple, undirected, weighted graph $G = (V,W,E)$ comprises a set of nodes or vertices $V = \{1,2,...,N\}$, a real-valued weight associated with each vertex $w_i \in W$ and a set of edges $E=\left\{(i,j) \text{ such that } i,j \in V, i \neq j\right\}$. Since the graph is undirected, we impose that $E$ is symmetric, i.e., $(i,j) \equiv (j,i)$. A clique $C \subseteq V$ is a subset of vertices such that any pair of nodes in $C$ shares an edge between them. The MVWCP is of finding a clique with maximum weight under inclusion, 
\begin{equation}
    C^* = \argmax_{\text{Clique } C} \sum_{i \in V} \mathbf{I}[i \in C] w_i \, . 
\end{equation}
We encode a clique as a bitstring $z$ of length $N$ where $z_i = 0/1$ indicates exclusion/inclusion within the clique.
With all unit weights, MVWCP reduces to the fundamental maxclique problem, among the original 21 NP-Complete problems identified by Karp.\cite{karp1975computational} Additionally, MVWCP is APX-Complete, meaning there is no known efficient algorithm to approximate MVWCP within a constant factor of its optimal objective value. Intuitively, these theoretical results bring to light the difficulty of MVWCP -- there is no efficient and tractable algorithm to solve it exactly. In addition, there is also no efficient algorithm to approximate its solution within a fixed relative error for any arbitrary graph instance.
\cite{gendreau1993solving}

\subsection*{QUBO Formulation Construction of MVWCP}

Applications of quantum algorithms to quadratic unconstrained binary optimization (QUBO) problems have been extensively studied in literature for a wide variety of combinatorial optimization problems. \cite{chapuis2017finding, li2010efficient}

For this problem, let us consider a graph \textit{G = (V, E)}, where the \textit{V} and \textit{E} are the vertex and edge sets, respectively. 
\\
Classically, QUBO \cite{lucas2014ising} problems are of the form
\begin{equation}
    \max_{z \in \{0,1\}^N} \sum_{i,j} A_{ij} z_i z_j + \sum_i B_i z_i
\end{equation}
Here, we briefly discuss the formulation of MVWCP as a QUBO problem. 
First, associate with each vertex $v_i \in V$ a decision-variable $x_i \in \{0,1\}$ indicating exclusion/inclusion in the solution clique.
We may frame a maximization objective for vertices in the clique as 
\begin{equation}
    \max \sum_i w_i x_i
\end{equation}
to recover the maximum-weighted solution clique. Next, consider two vertices $v_i, v_j$ which do not share an edge in $G$. Indeed, these vertices should not be both included in the clique. We may encode this constraint as
\begin{equation}
    x_i x_j  = 0 \quad \;\forall\, (i,j) \notin E 
\end{equation}

\subsection*{Pure Quadratic Formulation Construction of MVWCP}

We thus arrive at the constrained quadratic optimization problem
\begin{align}
    \max & \quad \sum_i w_i x_i \\
     \text{subject to } & \quad x_i x_j  = 0 \quad \;\forall\, (i,j) \notin E \\
     \text{with bounds } & \quad x_i \in \{0,1\} \quad \;\forall\, i \in V
\end{align}
We must convert this constrained problem into an unconstrained problem. 
It is a standard technique in the construction of optimization problems to add a penalty term for each violated constraint to the optimization objective as 
\begin{equation}
    \max \sum_i w_i x_i + \sum_{i,j} w_{ij} x_ix_j
\end{equation}
where $w_{ij} < 0$ if $(i,j) \notin E$ else $w_{ij} = 0$.

See that with this construction, for any pair of vertices included in the clique illegally,  the maximization objective is penalized by $w_ij$.
We set $w_{ij} = P < 0$ where $(i,j) \notin E$ to uniformly penalize all such inclusions.
With this, we may now form the quadratic unconstrained binary optimization problem as 
\begin{align}
    \max & \quad \sum_i w_i x_i + P \sum_{(i,j) \notin E} x_ix_j \\
     \text{with bounds } & \quad x_i \in \{0,1\} \quad \;\forall\, i \in V
\end{align}
We may form a cost Hamiltonian $H_C$ from this objective by the substitution $x_i = (1 - \sigma_i)/2$ to obtain. \cite{aharonov2008adiabatic}

\subsection*{Linear Formulation Construction of MVWCP}

\begin{align}
    \max & \quad \sum_i w_i x_i \\
    \text{subject to } & \quad x_i + x_j \leq 1 \;\forall\, (i,j) \notin E \\
    \text{with bounds } & \quad x_i \in \{0,1\}
\end{align}
This is an integer linear program. We can relax integrality to obtain 
\begin{align}
    \max & \quad  \sum_i w_i x_i \\
    \text{subject to } & \quad x_i + x_j \leq 1 \;\forall\, (i,j) \notin E \\
    \text{with bounds } & \quad x_i \in [0,1]
\end{align}
as the program to be fed into QAOA for warm-starting. \\

For reference, the following part of code is an example showing how to implement the linear formulation construction of the MVWCP. Initially, we consider the expected nodes and edges, as well as the epochs -a rounding parameter allowing QAOA to reach other states. After that, the constraints are being set for the checking the non-edge non-inclusion of the nodes. We are minimizing the weights by using scipy optimizer with a run of 200 iterations -for this case, which lasts approximately less than a few seconds. \\

\begin{lstlisting}[language=Python]
import numpy as np
from scipy import optimize

def solve_approximate_problem(nodes, edges, weights, eps = 0.25):
    non_edges = [(u,v) for u in nodes for v in nodes if u < v and (u,v) not in edges]
    
    constraint_mat = []
    for non_edge in non_edges:
        src, tgt = non_edge
        constraint = np.zeros(len(nodes))
        constraint[src] = 1
        constraint[tgt] = 1
        constraint_mat.append(constraint)

    constraint_mat = np.array(constraint_mat)
    b = np.ones(len(non_edges))
    weights = np.array(weights)

    weights *= -1

    res = optimize.linprog(
        weights, A_ub = constraint_mat, b_ub = b, bounds = [(0,1)] * len(nodes),
        options = {'maxiter': 200},
    )
    soln = res.x
    
    return np.clip(soln, eps, 1-eps)
\end{lstlisting}
State preparation involves single RY gate per qubit. Mixer has three gates per qubit, with one optimizable parameter. \cite{egger2021warm}

\break
\subsection*{QAOA and its variations}

The quantum approximate optimization algorithm (QAOA) is a hybrid classical-quantum algorithm used to solve combinatorial optimization problems. \cite{farhi2014quantum, pagano2019quantum, farhi2016quantum} 
QAOA is derived from the principles of quantum adiabatic computation, whereby starting $\ket{\psi_{\text{init}}}$ in the ground state of a some known simple Hamiltonian $H_M$ and evolving adiabatically in time finds the ground state of a more complex Hamiltonian $H_C$, $\ket{\psi_{\text{opt}}}$.

QAOA admits a alternating layer construction to its ansatz circuit, consisting of a cost-based operator $U_C(\vec\alpha)$ which encodes information about the problem structure, a mixing operator $U_M(\vec\beta)$ which allows the circuit to explore large swaths of the Hilbert space $\mathbb{C}^{2N}$ to find the ground state of $H_C$, and an optional counterdiabatic CD term $U_{CD}(\vec\gamma)$ which limits non-adiabatic evolution of the system. 
Each of these layers are complemented with a set of real-valued variational parameters $\vec{\alpha, \beta, \gamma}$ which are to be optimized to maximize the expected value $\expval{H_C}{\psi}$. 

QAOA proceeds by an iterative scheme of obtaining expected values of the Hamiltonian $H_C$'s energy with respect to a candidate solution $\ket{\psi_{\text{candidate}}}$ obtained from the trial circuit with $p$ layers with a certain set of parameters $(\vec{\alpha, \beta, \gamma})^{\times p}$ as
\begin{equation}
    E(\vec{\alpha}_1, \vec{\beta}_1, \vec{\gamma}_1, \ldots, \vec{\alpha}_p, \vec{\beta}_p, \vec{\gamma}_p) = \expval{H_C}{\psi_{\text{candidate}}} ; 
    \quad \quad 
    \ket{\psi_{\text{candidate}}} = \left[ \prod_{i=1}^p  U_C(\vec{\alpha}_i) U_M(\vec{\beta}_i) U_{CD}(\gamma_i)\right] \ket{\psi_{\text{init}}}
\end{equation}
The construction of this system allows for the fact that optimizing the parameters to the circuit recovers the ground state of the Hamiltonian $H_C$, which in turn recovers the optimal solution to the combinatorial optimization problem we seek to solve as 
\begin{align}
    E^* = E(\vec{\alpha}^*_1, \vec{\beta}^*_1, \vec{\gamma}^*_1, \ldots, \vec{\alpha}^*_p, \vec{\beta}^*_p, \vec{\gamma}^*_p) &= \argmax_{(\vec{\alpha, \beta, \gamma})^{\times p}}  \expval{H_C}{\psi_{\text{candidate}}} &\implies \\ 
    \ket{\psi_{\text{candidate}}} = \left[ \prod_{i=1}^p  U_C(\vec{\alpha}^*_i) U_M(\vec{\beta}^*_i) U_{CD}(\vec{\gamma}^*_i)\right] \ket{\psi_{\text{init}}}  &= \argmax_{\ket{\psi}}  \expval{H_C}{\psi} &\implies \\
    \ket{\psi_{\text{opt}}} = \argmax_{\ket{\psi}}  \expval{H_C}{\psi} & \approx \argmax_z \Cost(z) 
\end{align}
The optimal set of variational parameters is recovered through use of an classical optimization routine, such as the gradient-based BFGS (Broyden–Fletcher–Goldfarb–Shanno) optimizer and the gradient-free COBYLA (constrained optimization by linear approximation) optimizer. 
In this work, we consider three main types of QAOA ansatze:
\begin{enumerate}
    \item Conventional QAOA: This is the original QAOA ansatze proposed by Farhi et al., 2014. \cite{farhi2014quantum}  It takes the form of an alternating circuit with two non-commuting operators -- $U_C$ and $U_M$. 
    Here, the cost-operator $U_C = \exp(-i \alpha H_C)$ with $H_C$ being the cost Hamiltonian of the problem instance.
    The mixing operator $U_M = \exp(-i\beta H_M)$ with $H_M = \sum_{k=1}^N X_i$ applies a single parameterized X-rotation $RX(\beta)$ gate to each qubit. 
    This ansatze does not utilize a counterdiabatic term.
    Theoretical analysis of QAOA leads us to the result that as the number of layers in the circuit grows, the ground state obtained from QAOA when tends to the true solution of the combinatorial optimization problem, i.e.,
    \begin{equation}
        \lim_{p \to \infty} \ket{\psi_{\text{opt}}} \to \argmax_z \Cost(z)
    \end{equation}
\end{enumerate}
Many variations of QAOA have been proposed for NISQ implementation of the method, which are detailed in this excellent review paper \cite{blekos2024review}. 
Indeed, we prefer shallow circuits in light of imperfect error correction and noise in quantum device hardware, and research in this endeavor is broadly categorized into three broad directions: (a) modifying layer structure to allow for greater degrees of freedom, (b) design of specialized mixer operators through heuristic approaches based on analysis of the cost Hamiltonian $H_C$ and (c) intelligent warm-starting techniques for enhanced optimization capabilities, or reduced optimization costs for variational parameters. We investigate methodologies (a) and (c) through these QAOA ansatze:

\begin{enumerate}
    \item[2.] Counterdiabatic QAOA:  \cite{chandarana2022digitized}
    In qualitative terms, the principles of adiabatic computation, a large number of layers $p$ in the conventional QAOA ansatz plays a role of \emph{slowing} temporal evolution of the quantum state from $\ket{\psi_{\text{init}}}$ to $\ket{\psi_{\text{opt}}}$. Unfortunately, a large layer count with its associated gate count poses difficulty to near-term hardware. 
    Chandrarana et al. \cite{chandarana2022digitized} propose the digitized-counterdiabatic QAOA (DC-QAOA) ansatze, which involves the addition of a counterdiabatic operator $U_{CD}(-i \gamma H_{CD})$ to each layer, which reduces non-adiabatic evolution steps. 
    In our combinatorial optimization instance over a simple graph, the operator pool applicable to the problem is composed of various quantum-inspired algorithms, such as quantum approximate optimization algorithm (QAOA), variational quantum eigensolver (VQE), and other heuristic methods, \cite{yi2024iterative} the best of which may be heuristically determined. 
    We choose to only consider a CD term of the form $H_{CD} = \sum_{k=1}^N Y$, whereby the CD operation applied to each qubit is simply a Y-rotation $RY(\gamma)$. 
    Contrasting with the standard QAOA ansatz, this structure allows $Y$ rotations in addition to $X$ rotations from the mixer layer. Intuitively, this should lead to \emph{smoother}
    transitions in the quantum state evolution process.
    Indeed, this is a case of including additional degrees of freedom to each layer, in turn for a reduced layer count required for recovering good solutions. 

    \item[3.] Warm-starting QAOA: \cite{egger2021warm, truger2024warm, truger2022selection}
    Standard QAOA initializes the initial quantum state of the circuit to the uniform superposition state $\ket{+}^N$, i.e., the state where all bitstrings are equally likely. This is a reasonable starting state when no prior information of the solution of the system is known. However, with increasing dimension of problem, evolving $\ket{+}^{N}$ to $\ket{\psi_{\text{opt}}}$ is an computationally exhaustive process.
    
\end{enumerate}

\subsection*{Construction of Example Instances}

Here are more details on the formulation of the binding interaction graph (BIG) or docking graph, which we recreate here for completeness: we assume that pharmacophore points on both the protein -- $\{P_1, P_2, ..., P_n\}$ -- and ligand compound -- $\{L_1, L_2, ..., L_m\}$ -- have been selected through expert manual selection or data-driven approaches, and the distances of each pharmacophore pair are available to us. \cite{ding2024molecular}
We construct pairs of pharmacophores based on their possibility to bind, abstracted into a vertex as $v_p = (P_p, L_{p'})$, which can have at most $\frac{N(N-1)}{2}$ edges between them.
Any two vertices $v_p = (P_p, L_{p'}), v_q = (P_q, L_{q'})$ share an edge between them if $\dist(P_p, P_{q}) + \dist(L_{p'}, L_{q'}) \le \delta$ where the threshold $\delta$ is indicative of the interaction distance between pharmacophores and flexibility of the ligand-protein ensemble in its comformational analysis. 
In this study, we consider graphs of various sizes derived from chemical data. To test hypotheses and more complete experiments on easily interpretable graphs, we supplement these \emph{real} instances with custom, aphysical instances. We describe these in the following.

Herein we take the cocrystal structure of Helios and E3 ubiquitin ligase degrader complex as an example to illustrate how to convert molecular docking into a clique-finding problem. The zinc-finger transcription factor Helios plays a critical role in suppressive activity of regulatory T cells (Tregs) and thus is an attractive target for immuno-oncology.\cite{wang2021acute} Given the low druggability of Helios, degradation of Helios has a high potential in such therapeutics. Figure 1a shows the co-crystal structure of Helios with ligand ALV1, which recruits the E3 ubiquitin ligase substract receptor cereblon (CRBN). Figure 1b indicates key pharmacophores identified from the binding site, located at the interface between Helios and CRBN. The pocket pharmacohpores feature two hydrophobic phc4s (P1 and P4) at the two ends of the pocket. P1 is the center of three trptophines to the left and P4 is formed by a histidine and proline at another side. These two major hydrophbic phc4s are cooperated with other polar interaction phc4s (P2, P3, P5) to mediate the interface between the ligand and zinc-finger (e.g. P5 defines the hydrogen bond interaction to the backbone carbonyl). 

An effective ligand should satisfy those pharmacophores and engage a tight interaction with the pocket. Figure 1c defines pharmaophres for a binding ligand. L1, L6, and L9 indicate the hydrophobic area of the ligand, while other L\textsubscript{i} present polar phc4 such as hydrogen bond donor and hydrogen bond acceptor. Next, we define how phc4s from the ligand should interact with pocket phc4s in a clique. Different from Ding {\it et al}'s method\cite{ding2024molecular} in which a weighted matrix is used to define how likely each ligand phc4s can overylap with each pocket phc4s, we simplify the problem by only allowing certain type of phc4 to interact, {\it i.e.}, only the same type of phc4s can overlap in a clique (Figure 1d). Matched phc4s (1s in Figure 1d matrix) are employed to construct nodes in the binding interaction graph as depicted in Figure 1f, in which the number of the nodes matches the number of 1s in Figure 1d. This method can significantly reduce the number of nodes in the interaction graph, and make it possible to apply the QAOA method to more complex and practical systems. 

We then follow strategies by Ding {\it et al} \cite{ding2024molecular} and Banchi {\it et al} \cite{banchi2020molecular} to construct edges connecting those nodes. Each node indicates contact between a ligand phc4 $L_i$ and pocket phc4 $P_i$. As in Figure 1e, for two nodes containing $L_i$, $P_i$, $L_j$, and $P_j$, if the distance between $L_i$ and $L_j$ ($d_l$) is not different from the distance between $P_i$ and $P_j$ ($d_p$) by a predefined buffer value ($2\tau$), then there is an edge connecting the two nodes. The physical meaning of an edge in the binding interaction graph is the two phc4 contacts matching the ending nodes of that edge can co-exist in the binding pose. All phc4 contacts in a binding pose thus form a fully connected sub-graph as depicted in orange in Figure 1f. Therefore, we convert the molecular docking problem into a graph theory problem of how to find the maximum fully-connected sub-graph. 
\begin{figure}
    \centering
    \includegraphics[width=0.7\linewidth]{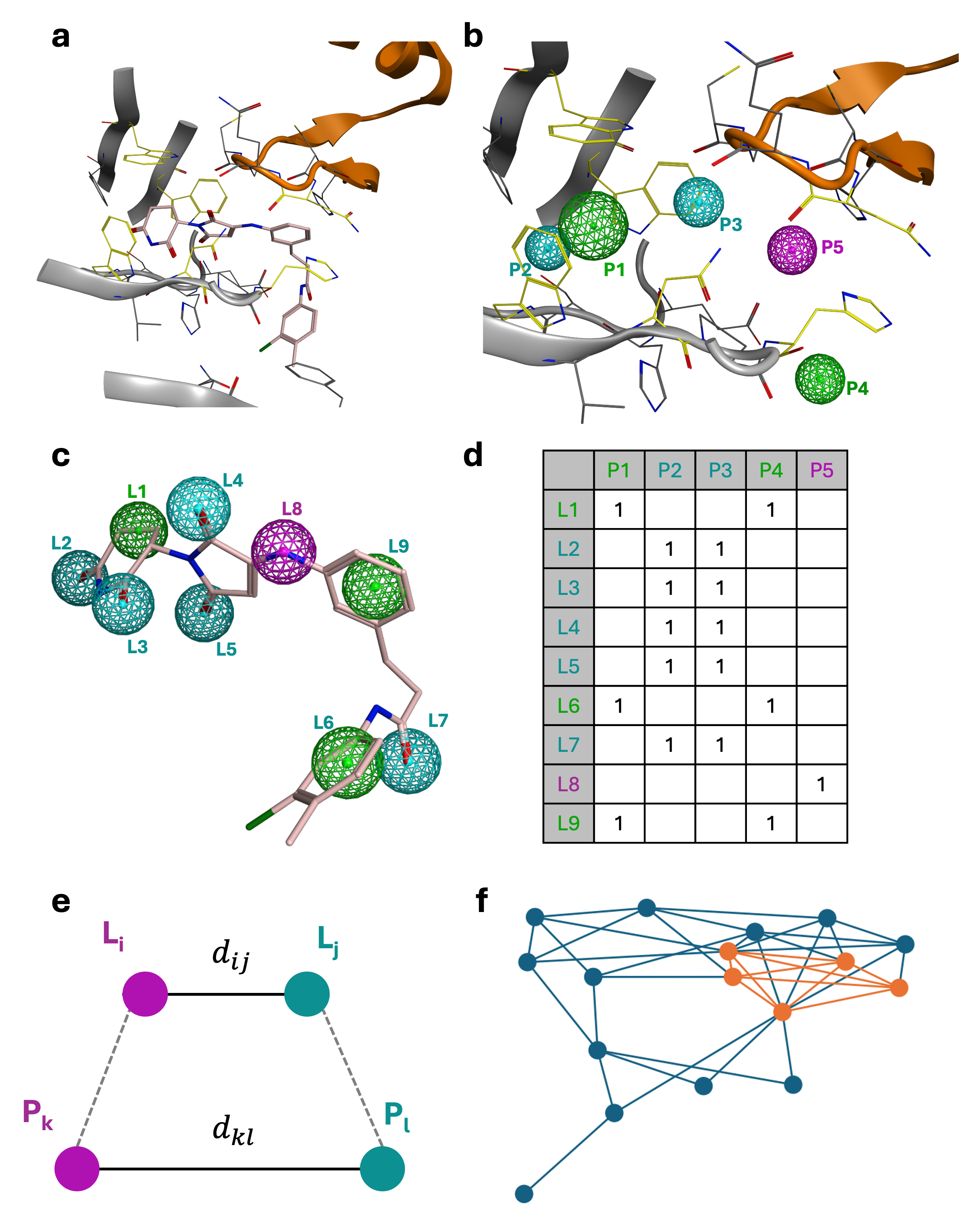}
    \caption{Illustration of converting molecular docking to clique finding problem. a. Crystal structure of protein-ligand binding complex \cite{ding2024molecular}}
    \label{fig:enter-label}
\end{figure}
\FloatBarrier

\section*{Results}
\label{results}

\subsection*{NVIDIA CUDA-Q}
This work was done with CUDA-Q, an open-source, QPU-agnostic platform designed for accelerated quantum supercomputing. CUDA-Q provides a hybrid programming model enabling seamless scaling between CPU, GPU, multi-GPU and QPU platforms. 

Using CUDA-Q, we were able to use the built-in parallelization features which enable distribution of quantum workloads across the available architectures. For example, one can calculate expectation values of a Hamiltonian with respect to a quantum kernel with: 

\begin{verbatim} cudaq.observe(kernel, hamiltonian). \end{verbatim}
This execution function can be parallelized with respect to any of the input arguments it accepts:
\begin{verbatim} 
cudaq.observe_async(kernel_1, hamiltonian, qpu_id = 0)
cudaq.observe_async(kernel_2, hamiltonian, qpu_id = 1)
\end{verbatim}
where kernel\_1 and kernel\_2 are circuits that compute the ansatz circuit that is used in the code and the qpu\_id indexes the architecture used to execute the quantum circuit. In our case, we parallelize the circuit execution on A100 GPUs but as quantum hardware matures, these qpu\_ids will refer to distributed quantum systems.

\subsection*{Computational Results - Experiments}

This section presents the results of the different experiment runs were executed in Pfizer's on premises clusters.   Background on the results, the impediments, the fine tuning parameter's setting, and the different QAOA approaches are given.

For a more comprehensive understanding on the results, computational resources used are presented. Methods and approaches were implemented in Python and the code was executed, as mentioned before, in Pfizer's High Performance Computing (HPC) environment and specifically in A100 GPUs. Each of those GPU has 40GB of memory and one GPU was used for every single run. GPUs are well-suited for running QAOA due to their inherent parallelism and computational efficiency.

Unlike traditional CPUs, which are optimized for sequential processing, GPUs can execute thousands of operations simultaneously, making them an ideal tool to highly parallel nature of quantum algorithms. In the context of molecular docking, where large-scale simulations and optimizations are required, leveraging GPUs can significantly reduce computation time and enhance the performance of QAOA. This paper explores the practical aspects of implementing QAOA on GPUs, including optimization strategies and integration with existing molecular docking frameworks.

Warm-starting technique was used as an initial step while executing our experiments. Warm-starting improves the efficiency of optimization algorithms, including QAOA. It involves initializing the quantum algorithm with a solution that is close to optimal, often obtained from a classical algorithm or a previous iteration of the quantum algorithm itself. Starting by a near to optimal solution, the algorithm can converge more quickly to the final optimized one, reducing the number of quantum operations required. In the NISQ era, where quantum resources are limited, warm-starting can be particularly beneficial, as it helps mitigate the effects of noise and decoherence in quantum devices. Benefits of warm-starting in the context of QAOA for molecular docking, especially when running on GPU platforms are examined at this effort.

After this initial step, COBYQA (constrained optimization by quadratic approximation) was used  as our main computational and optimization method. Other methods were used as well, like RCD, COBYLA but with not so promising results such as COBYQA. Using COBYQA combining to running the algorithm in GPUs, better running times and convergence observed.

\subsubsection*{Results on 14-qubits}
The computational results for the 14-qubit instance were obtained using the COBYQA  algorithm. In this approach, we employed a penalty parameter of $P = 1$ and utilized a single layer, $n = 1$ in the optimization process.
Additionally, we conducted a comparative analysis of the algorithm's performance with and without the application of a warm-start technique. For the specific analysis, a quadratic classical solver were used to derive the initial solution for the optimization process.
\begin{figure}
    \centering
    \includegraphics[width=1.0\linewidth]{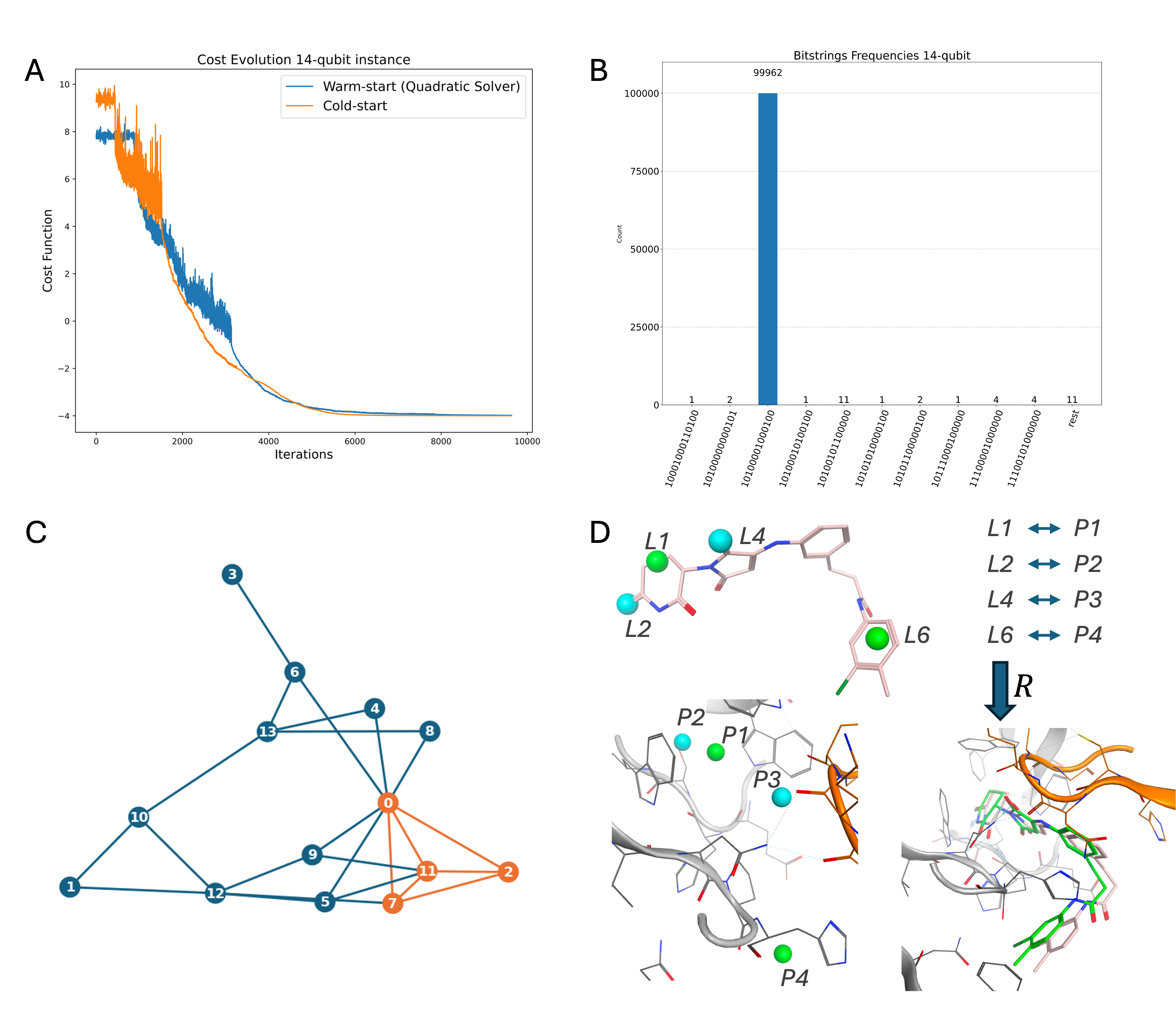}
    \caption{A. Cost evolution comparison of applying a warm-start technique and not. The warm-start technique provides a better initial convergence point for the optimization process. For this instance example the convergence time is almost the same. B. Histogram of the top 10 sampled frequencies for 14-qubit instance. The most sampled bitstring correspond to the actual ground-truth state. C. Graph of 14-qubit instance with the solution colored.}
    \label{fig:enter-label}
\end{figure}
\FloatBarrier

The figure above illustrates that the warm-start technique facilitates the optimization algorithm by providing a more favorable initial convergence state. Furthermore, it effectively reduces the amplitude of oscillations, especially for the first iterations, compared to runs without a warm-start. However, for this specific instance the convergence time remains approximately the same.  

The histogram below presents the distribution of bistring solutions from the above experiments. The most frequently sampled solution exhibits a well-defined structure and it strongly aligned with the ground-truth state of the specific 14-qubit instance. The distribution reflecting the dominance of a particular solution.
Furthermore, by analyzing the graph Figure 4., we can validate the results, as it confirms the presence of a single 4-node clique, which corresponds to the maximum weighted clique in the graph.

\subsubsection*{Results on 17-qubits}
Below we can see different convergence times for three different scenarios. One without a warm-start, one with a warm-start using a linear classical solver and one with warm-start using quadratic classical solver. 
As seen in Figure 5 (a), for warm-start with linear solver, we are using different number of layers and different number of penalties. The number of layers are $n = 1$, $n = 2$. Number of penalties are $P=1$ and $P=2$. For each number of layer ($n$), one can notice that for $n = 1$ the convergence is done quicker in a smaller amount of iterations ( $\sim$ 2000 iterations in this case). When $n$ is being increased, the algorithm needs more time and iterations to converge and to get to the expected solution. The same behavior is being observed goes while $P$ being equal to 2, which is more hard to stabilize and converge to the required result (number of layers in this case $n=1$). 
For warm-start with quadratic solver, Figure 5 (b),  same configuration is used, for each different experiment/run, different number of layers are used with different penalties ($P = 1$, $P=2$). For $n = 1$, we can see that convergence is quicker in this case as well, but after 2000-3000 iterations it stabilizes. In $n = 2$ and case (penalty stays $P=1$), we can observe that the convergence needs time to consolidate and probably increase the number of iterations. For $P=2$ and $n=1$ case, we can see that convergence is not good. 
 By increasing the penalty $P$, we can observe that it does not help getting good results. On the opposite, when we increase the number of layers $n$ better results are being returned. 

\begin{figure}[ht]
\begin{subfigure}{0.33\textwidth}
    \centering
        \includegraphics[width=1\linewidth]{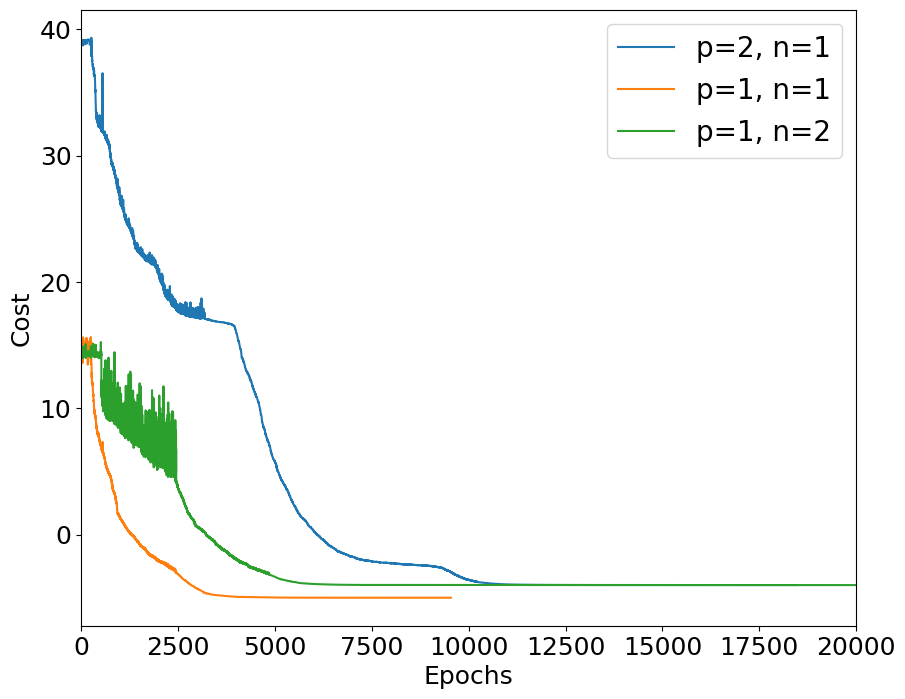}
    \caption{Linear warm-start (17-qubits)}
    \label{enter-label:sfig1}
\end{subfigure}
\hfill %
\begin{subfigure}{0.33\textwidth}
    \centering
    \includegraphics[width=1\linewidth]{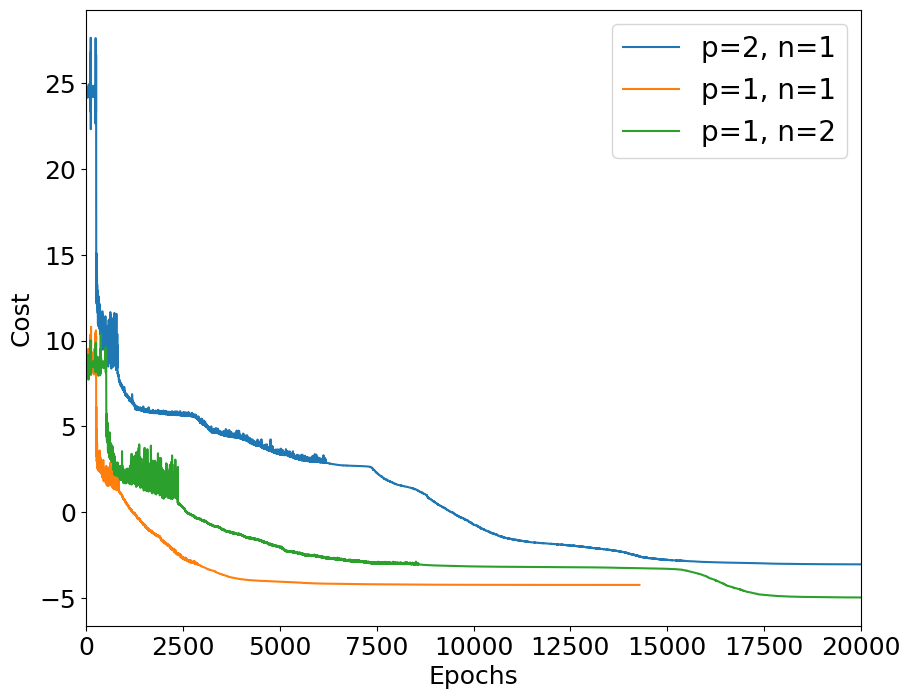}
    \caption{Quadratic warm-start (17-qubits)}
    \label{enter-label:sfig2}
\end{subfigure}
\hfill %
\begin{subfigure}{0.33\textwidth}
    \centering
    \includegraphics[width=1\linewidth]{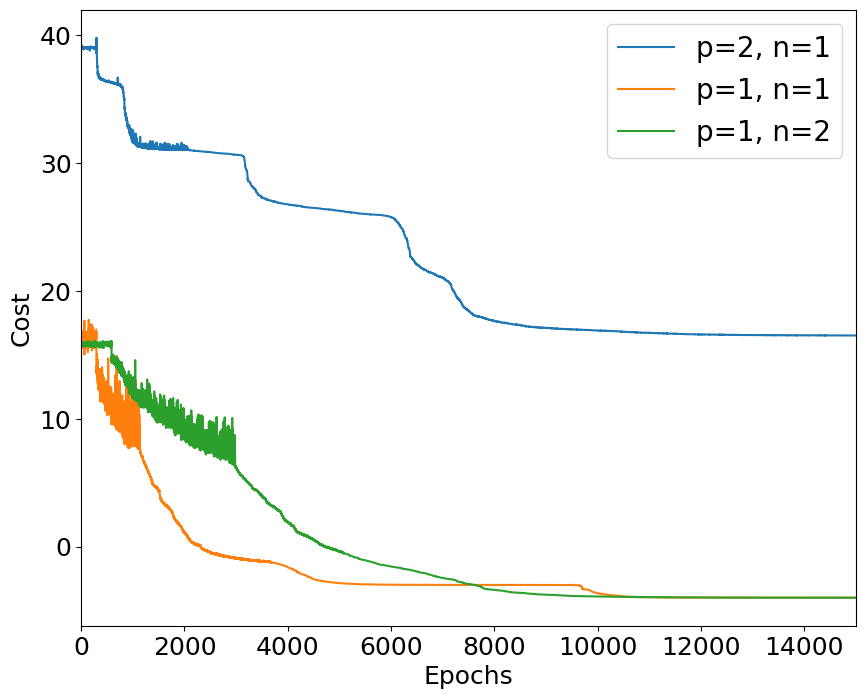}
    \caption{Cold-start (17-qubits)}
    \label{enter-label:sfig3}
\end{subfigure}
    \caption{In Panel (a), (b) and (c) we illustrate the cost evolution determined by DC-QAOA approach by using COBYQA with linear warm-start, quadratic warm-start and cold-start. For each case, we generated and optimized with varying penalties ( $P=1$, $P=2$), varying layers ($n = 1$, $n = 2$) and 40000 iterations. }
    \label{fig:enter-label}
\end{figure}
In Figure 5 (c), and for cold-start  method, we can observe that the number of iterations is increased to reach convergence, but again with not so good convergence. Same configuration is applied here for each different experiment/run. After increasing $n$, from $n = 1$ to $n = 2$, we observe that more iterations are needed to reach convergence. For $P=2$ and $n=1$, the cost is higher as one can see in the plot and convergence are not meeting the expected result, hence not getting the proper solution. 

In Figure 6 (a), (b) and (c), we compare all these separate methods by using the cost functions of each one. The configuration is the same, with  number of layers $n = 1$ and penalty $P = 1$, $n=2$ and  $P=1$, $n=1$ and $P=2$. For first case, it seems to converge and stabilize after some number of iterations, we are getting results for linear and quadratic but we are not getting a bitstring as a result that corresponds to the actual pharmacophores for the cold-start case. For the second case,  convergence seems to be good but only quadratic method is giving a solution, with the other two not giving any. In the last case, increasing the penalty to 2 and having number of layers equal to 1, all three methods are not giving any solution. 
\begin{figure}[ht]
\begin{subfigure}{0.33\textwidth}
    \centering
    \includegraphics[width=1\linewidth]{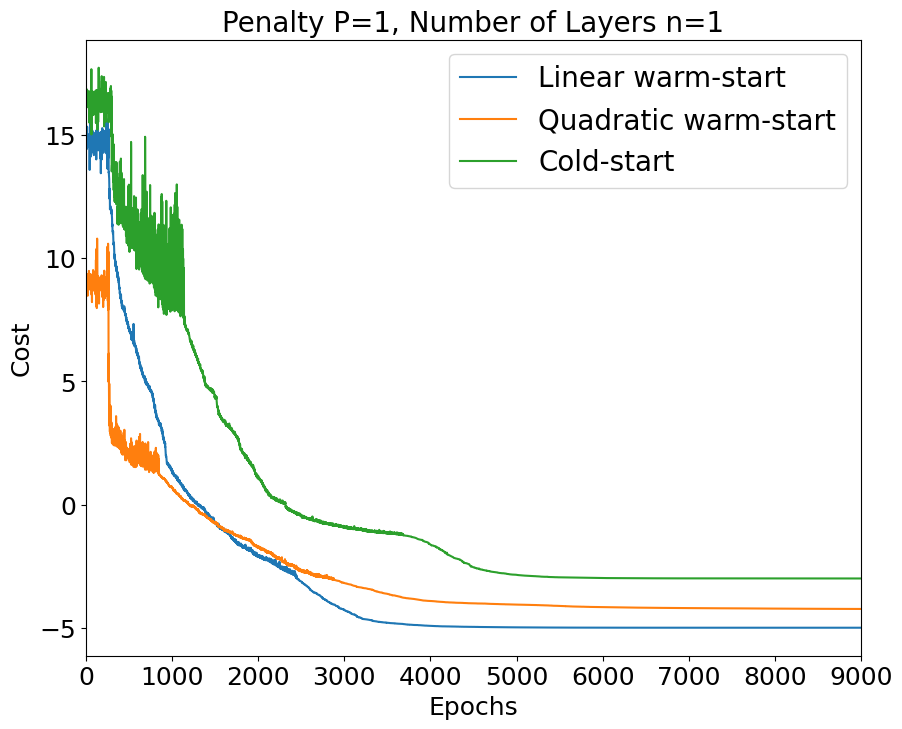}
    \caption{$P=1$, $ n=1$}
    \label{enter-label:sfig1}
\end{subfigure}
\hfill %
\begin{subfigure}{0.33\textwidth}
    \centering
    \includegraphics[width=1\linewidth]{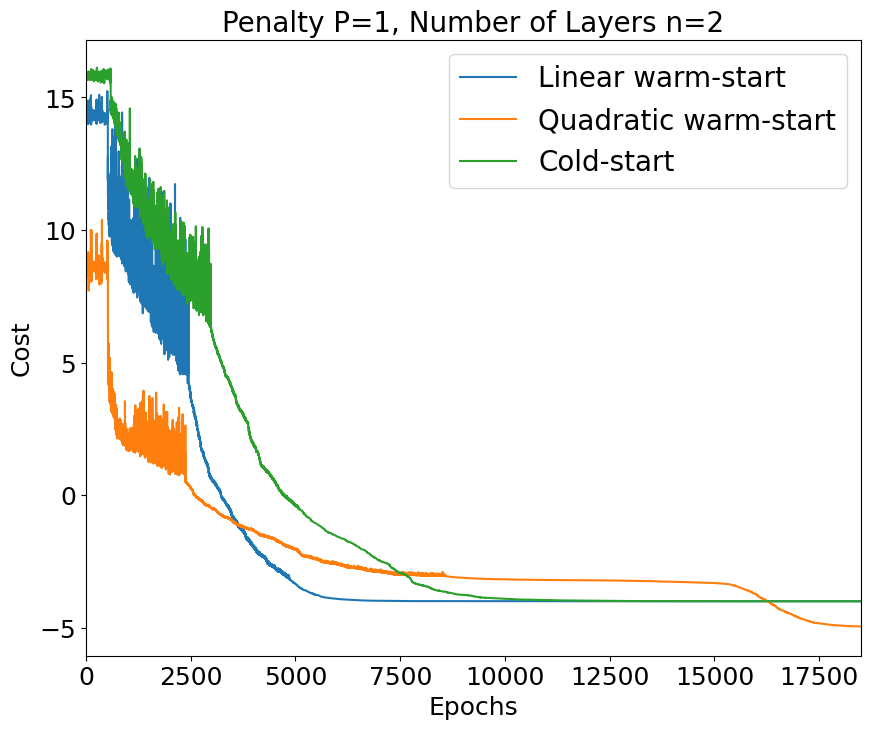}
    \caption{$P=1$, $n=2$}
    \label{enter-label:sfig2}
\end{subfigure}
\hfill %
\begin{subfigure}{0.33\textwidth}
    \centering
    \includegraphics[width=1\linewidth]{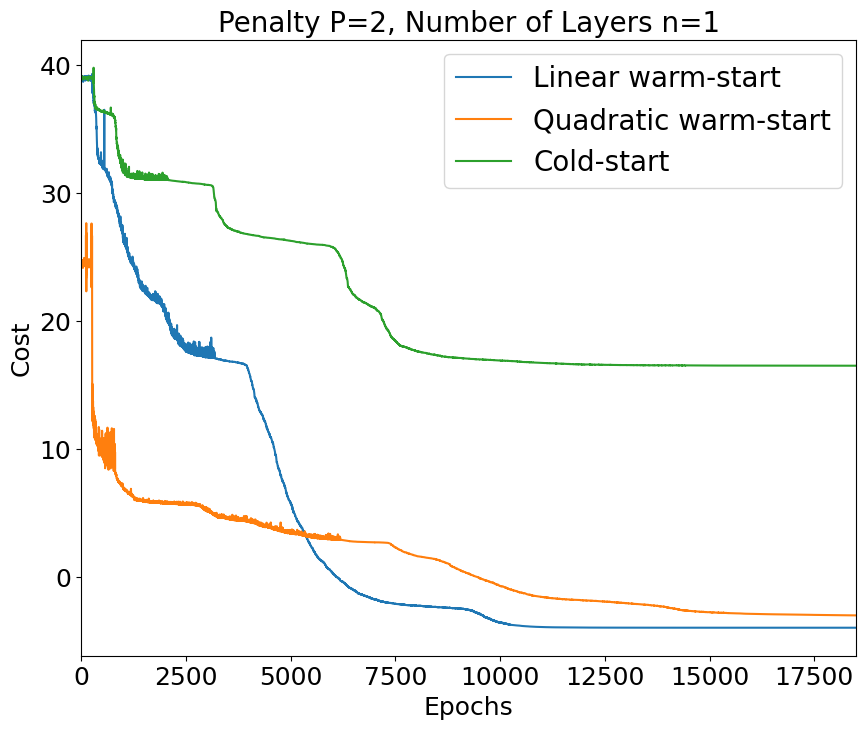}
    \caption{$P=2$, $n=1$}
    \label{enter-label:sfig3}
\end{subfigure}
\caption{Different method runs (quadratic, linear and cold-start), compared to each other, with varying layers and penalties. All of these are COBYQA optimization runs.}
    \label{fig:enter-label}
\end{figure}
\FloatBarrier

From the bitstring solution histogram, we can observe that the most sampled solution is clean and it corresponds to the actual  ground-truth of the instance. The top ten of each experiment is sampled and the one that has the highest cost (and corresponds to the ground-truth) is the bitstring 10100001000100100. Each "1" represents an edge and according this bitstring the edges are: [0,2,7,11,14].  This can be proven and seen in Figure 8, where we can see the colored edges which are creating the max clique and represent as well the pharmacophores of the solution. 

\begin{figure} [ht]
    \centering
    \includegraphics[width=0.5\linewidth]{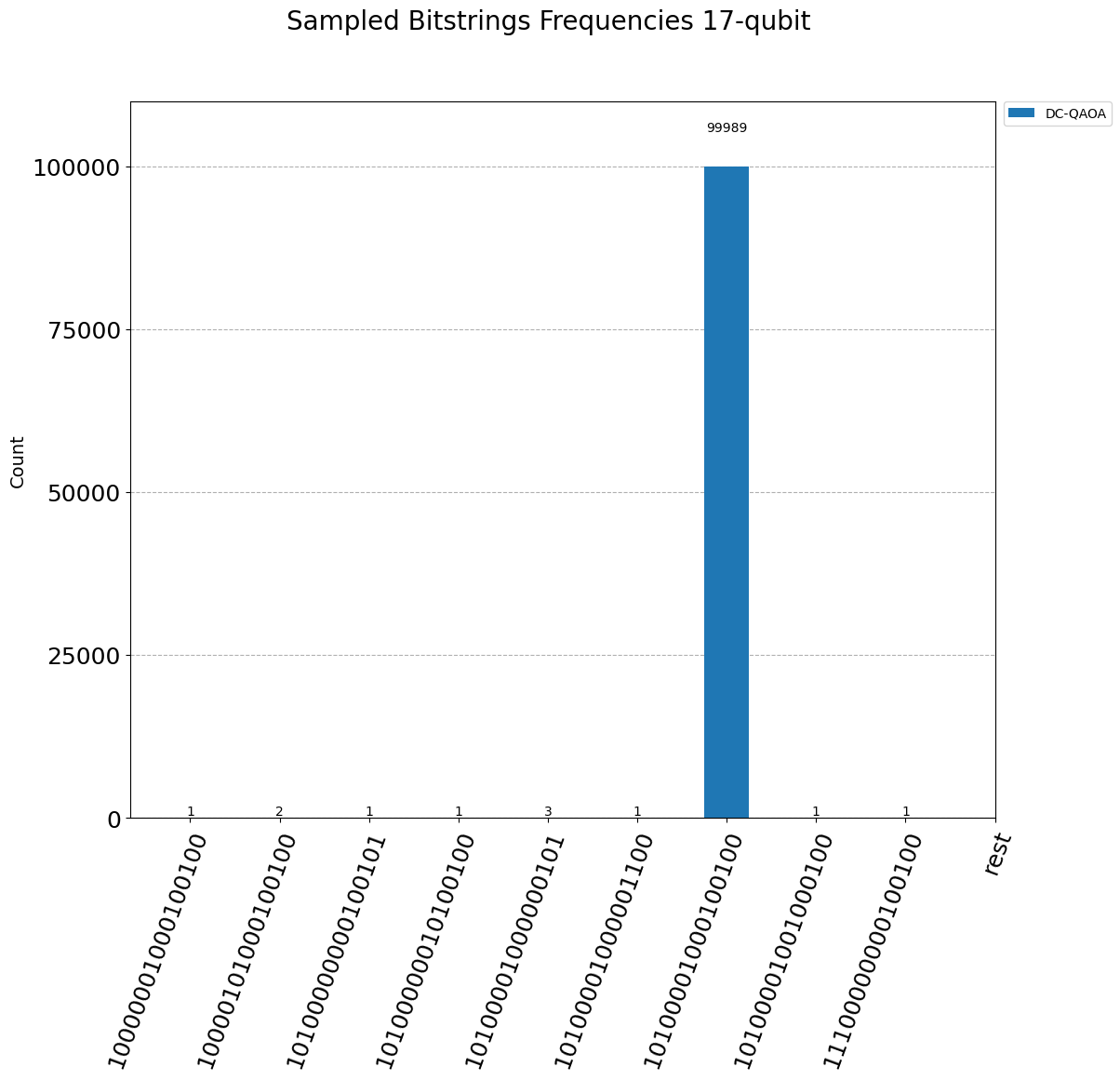}
    \caption{Sampled frequencies by bitstrings for 17-qubits with solution of pharmacophores. We are selecting the top 10 of each experiment run, where each bitstring with the highest cost, represents the solution.  }
    \label{fig:enter-label}
\end{figure}
\FloatBarrier
\begin{figure}
    \centering
    \includegraphics[width=0.5\linewidth]{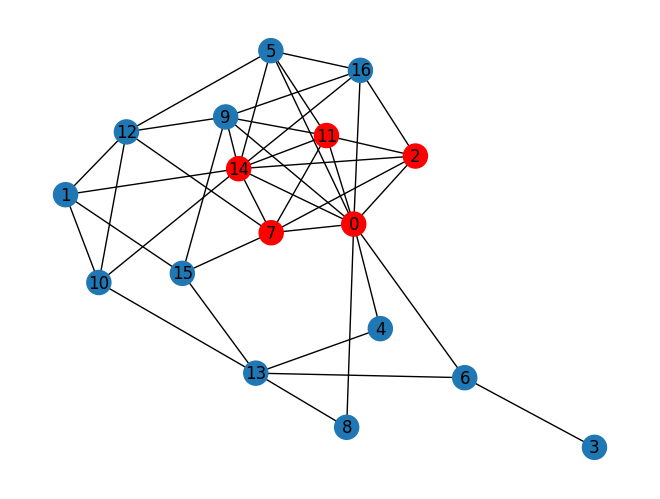}
    \caption{Graph for 17-qubit system which present the solution. This graph with colored (red) edges represents the pharmacophores of the solution.  }
    \label{fig:enter-label}
\end{figure}
\FloatBarrier

\subsection*{Discussion and Conclusions}
\label{conclusion}

In this study, we explored the potential of the quantum approximate optimization algorithm (QAOA) for addressing real molecular docking problems. The results indicate that this approach can enhance the accuracy and efficiency of docking simulations and apply it to real world docking problems and data.
Our findings suggest that tensor networks and multi-GPU workflows \cite{CUDA-Q} hold promise for future research, and we plan to explore this area further in subsequent papers.
We want also to apply QAOA to real docking problems, focusing on building the binding interaction graph differently and benchmarking it on even larger sets of protein-ligand complexes.

Looking ahead, we aim to transfer these computations to real quantum processing units (QPUs),  - if possible. This step will allow us to validate our findings in a practical quantum computing environment and potentially unlock new capabilities for solving complex docking problems. Another step, is to extend this use case and "stress" the solution by feeding more qubits, specifically for 20-qubits and then increase to 34-qubits. Increasing the qubits will reveal the algorithm's limitations and allow benchmarking on larger, more complex sets of protein-ligand complexes.

\section*{Acknowledgements}
This research was conducted as a collaboration between Nvidia and 
Pfizer.
The authors would like to acknowledge from Pfizer side, Xinjun Hou and Joy Yang for their ongoing support on the research. Also, we would like to acknowledge from Nvidia collaboration Reynaldo Gomez, Marwa H. Farag and Zohim Chandani for their active guidance, support on CUDA-Q platform and continuous effort of reviewing our methods and computational results.

 \bibliography{mybib.bib}
  \nocite{*}





\end{document}